\documentclass[epj]{svjour}

\usepackage{amssymb}
\usepackage{graphicx}

\title{Interaction effects on atomic laboratory trapped Bose-Einstein condensates}

\author{Fabio Briscese \inst{1}  \inst{2}}

\institute{Istituto Nazionale di Alta Matematica Francesco Severi,
Gruppo Nazionale di Fisica Matematica, Citt\`a Universitaria, P.le
A. Moro 5, 00185 Rome, EU. \and Dipartimento SBAI, Sezione di
Matematica, Sapienza Universit\`a di Roma, Via Antonio Scarpa 16,
00161 Rome, EU. \\
\email{fabio.briscese@sbai.uniroma1.it} }

\abstract{We discuss the effect of inter-atoms interactions on the
condensation temperature $T_c$ of an atomic laboratory trapped
Bose-Einstein condensate. We show that, in the mean-field
Hartree-Fock  and semiclassical approximations, interactions
produce a shift $\Delta T_{c}/T_{c}^{0} \approx b_1
(a/\lambda_{T_c}) + b_2 (a/\lambda_{T_c})^2 +
\psi\left[a/\lambda_{T_c}\right]$ with $a$ the s-wave scattering
length, $\lambda_T$ the thermal wavelength and
$\psi\left[a/\lambda_{T_c}\right]$ a  non-analytic function such
that $\psi\left[0\right] = \psi'\left[0\right] =
\psi''\left[0\right] = 0$  and $|\psi'''\left[0\right]| = \infty$.
Therefore, with no more assumptions than Hartree-Fock and
semiclassical approximations, interaction effecs are perturbative
to second order in $a/\lambda_{T_c}$ and the expected
non-perturbativity of physical quantities at critical temperature
appears only to third  order. We compare this finding with
different results by other authors, which are based on more than
the Hartree-Fock and semiclassical approximations. Moreover, we
obtain an analytical estimation for $b_2 \simeq 18.8$ which
improves a previous numerical result. We also discuss how the
discrepancy between $b_2$ and the empirical value of $b_2 = 46 \pm
5$ may be explained  with no need to resort to beyond-mean field
effects. }

\begin{document}

\maketitle

Bose-Einstein condensates (BECs) are  produced in the laboratory
in laser-cooled, magnetically-trapped ultra-cold bosonic clouds of
defferent atomic species, e.g. $_{37}^{87}Rb$ \cite{Ander},$\
_{3}^{7}Li$ \cite{Bradley}, $ _{11}^{23}Na$ \cite{Davis},
$_{1}^{1}H$ \cite{Fried}, $_{37}^{87}Rb$ \cite {Cornish},
$_{2}^{4}He$ \cite{Pereira}, $_{19}^{41}K$ \cite{Mondugno}, $
_{55}^{133}Cs$ \cite{Grimm}, $_{70}^{174}{Yb}$ \cite{Takasu03} and
$ _{24}^{52}Cr$ \cite{Griesmaier}, and represent a relevant test
for fundamental quantum theory, see Ref.s
\cite{dalfovo,parkinswalls,burnett} for a review. Moreover, it has
recently shown that BECs can be used  to constrain Quantum Gravity
models \cite{briscese}.

Inter-particle interactions play a fundamental role in the physics
of condensation since they are necessary to reach thermal
equilibrium. The effect of interactions on the condensation
temperature $T_c$ of a BEC has been the subject of  extensive
research from first works of Lee and Yang \cite{a1,a2} until more
recent works \cite{a3,a4,a5,a6,a7,a8,a9,a10,a11,a12}. The first
effort has been devoted to the study of the condensation
temperature of interacting uniform BECs. In this case,
interactions are irrelevant in the mean field (MF) approximation,
see \cite{a7,a10,a11,a12} for reviews. However, interactions
produce a shift in the condensation temperature of uniform BECs
with respect to the ideal non interacting case, which is due to
beyond-MF effects related to quantum correlations between bosons
near the critical point. This effect has been finally quantified
in Ref.s \cite{a7,a8} as

\begin{equation}
\Delta T_{c}/T_{c}^{0}\simeq \, 1.8 \, (a/\lambda_{T^0_c})
\label{deltaTsuTUNIFORM}
\end{equation}
where $\Delta T_{c}\equiv T_{c}-T_{c}^{0}$ with $\,T_{c}$ the
critical temperature of the gas of interacting bosons, $T_{c}^{0}$
the condensation temperature in the ideal non-interacting case,
$\lambda_{T^0_c}\equiv \sqrt{2 \pi \hbar^2/(m k_B T^0_c)}$ the
thermal wavelength at temperature $T^0_c$, $n$ the boson number
density and $a$  the s-wave scattering length used to parameterize
inter-particle interactions \cite{dalfovo,parkinswalls,burnett}.
Hereafter we limit our attention to repulsive interactions with $a
\geq 0$. Relation (\ref{deltaTsuTUNIFORM}) implies that the shift
in the condensation temperature of a uniform gas due to beyond-MF
effects is positive for repulsive interactions.

Laboratory condensates are not uniform BECs since they are
produced in atomic clouds confined in magnetic traps, but they can
be described in terms of harmonically-trapped BECs consisting of a
system of $N$ bosons trapped in an external spherically symmetric
harmonic potential $V = m \omega^2 x^2/2$, with $\omega$ the
frequency of the trap and $m$ the mass of the bosons.

Neglecting inter-boson interactions,  the condensation temperature
is $k_B T^0_c = \hbar \omega \left(N/\zeta(3)\right)^{1/3}$
 \cite{dalfovo}. However, when interactions are considered one
finds a shift in $T_c$ given by

\begin{equation}
\Delta T_{c}/T_{c}^{0}\simeq b_1 (a/\lambda_{T^0_c}) + b_2
(a/\lambda_{T^0_c})^2. \label{deltaTsuNONTUNIFORM1}
\end{equation}
with $b_1 \simeq -3.426$ \cite{dalfovo} and $b_2 \simeq 11.7$
\cite{condensatePRL2}, implying that  $\Delta T_{c}$ is negative
for repulsive interactions. The result in
(\ref{deltaTsuNONTUNIFORM1}) is in excellent agreement with
laboratory measurements of $\Delta T_{c}/T_{c}^{0}$
\cite{b1,b2,b3,condensatePRL} to first order in
$(a/\lambda_{T^0_c})$ but shows some tension with data to second
order $(a/\lambda_{T^0_c})^2$. In Ref.\cite{condensatePRL}, high
precision measurements of the condensation temperature of $^{39}K$
in the range of parameters $N\simeq (2-8) \times 10^5$, $\omega
\simeq (75-85) Hz$, $10^{-3} < a/\lambda_{T^0_c} < 6 \times
10^{-2}$ and $T_c \simeq (180-330) nK$ have detected second order
effects in $\Delta T_{c}/T_{c}^{0}$. The measured $\Delta
T_{c}/T_{c}^{0}$ is well fitted by the quadratic polynomial
(\ref{deltaTsuNONTUNIFORM1}) with best-fit parameters $b_1^{exp}
\simeq -3.5 \pm 0.3$ and $b_2^{exp} \simeq 46 \pm 5$ so that the
value $b_2\simeq 11.7$ \cite{condensatePRL2} is strongly excluded
by data.

The discrepancy between (\ref{deltaTsuNONTUNIFORM1}) and data may
be due to beyond-MF effects (see Ref.\cite{condensatePRL2}). We
mention that beyond-MF effects are expected to be important near
criticality, where the physics is often non-perturbative. It is
therefore reasonable that a beyond-MF treatment may give a correct
estimation of $b_2$. However, this is not proven since beyond-MF
effects have been calculated in the case of uniform condensates
\cite{a7,a8} but they are still not clearly understood for trapped
BECs \cite{c1,c2,c3,c4,c5}. Thus it seems that  it is currently
not possible to ascertained whether  the discrepancy between $b_2$
and $b_2^{exp}$ can be explained in the MF context or  is due to
beyond-MF effects.

In this work we discuss how  this problem may be resolved within a
MF treatment. We assume that the MF Hatree-Fock approximation and
the semiclassical limit are both valid and, with no more
assumptions, we show that the temperature shift $ \Delta
T_{c}/T_{c}^{0}$ is a non analytic function of the s-wave
scattering length  $a$ at $a=0$ and has the following asymptotic
expansion for small $a \ll \lambda_{T^0_c}$

\begin{equation}
\Delta T_{c}/T_{c}^{0} \approx b_1 (a/\lambda_{T^0_c}) + b_2
(a/\lambda_{T^0_c})^2 +  \psi\left[a/\lambda_{T^0_c}\right]
\label{deltaTsuT Asymptotic}
\end{equation}
with  $\psi\left[a/\lambda_{T^0_c}\right]$ a  non analytic
function such that $\psi\left[0\right] = \psi'\left[0\right] =
\psi''\left[0\right] = 0$   and $|\psi'''\left[0\right]| =
\infty$. Thus, we obtain a perturbative result to second order in
$a/\lambda_{T^0_c}$ and we show that non-perturbativity appears
only to third order. As we will discuss, this is in contrast with
different findings based on more assumtions than the Hartree-Fock
and semiclassical limits \cite{c1} in which non-perturbativity of
$\Delta T_{c}/T_{c}^{0}$ emerges to second order in
$a/\lambda_{T^0_c}$.

We stress that the additional non-perturbative term
$\psi\left[a/\lambda_{T^0_c}\right]$ in (\ref{deltaTsuT
Asymptotic}) may explain the discrepancy between
(\ref{deltaTsuNONTUNIFORM1}) and the data with no need to resort
to beyond-MF effects. However the problem of the exact
determination of $\psi\left[a/\lambda_{T^0_c}\right]$ goes beyond
the intent of this letter and will be addressed elsewhere.

We obtain an  estimation of  $b_2 \simeq 18.8$ in the MF
approximation  which  improves upon  the value $11.7$ obtained in
\cite{condensatePRL2}. Such a difference may be due to the fact
that in \cite{condensatePRL2} the parameter $b_2$ is estimated
numerically while we obtain an analytic result. Therefore, the two
results are compatible within the limits of precision of numerical
calculations in \cite{condensatePRL2}.

In what follows we first introduce the MF Hartree-Fock and
semiclassical approximations used in this work. Then we  describe
the procedure used to calculate $b_1$. Subsequently we proceed to
calculate the coefficient $b_2 \simeq 18.8$ and compare the result
with the numerical estimation obtained in \cite{condensatePRL}.
Afterwards we show that $\partial^3_a\Delta T_{c}$ diverges at $a
=0$ so that the function $\Delta T_{c}$ is not analytical there.
Finally, we discuss the consequences of this finding and argue how
the discrepancy of (\ref{deltaTsuNONTUNIFORM1}) with data can be
interpreted in the MF-framework.

In the MF Hartree-Fock approximation, bosons are treated as a non
interacting gas that experiences a MF interaction potential
$\propto g \, n(x,g)$, where $g = (4 \pi \hbar^2 a/m )$
\cite{dalfovo,parkinswalls,burnett} and $n(x,g)$ is the density of
bosons at the point $x$ which also depends on $g$, see for
instance Eq.(\ref{semiclassicaln2}) below, so that the
Hartree-Fock hamiltonian is
\begin{equation}\label{HF Hamiltonian}
H_{HF} = \frac{P^2}{2m} + V(x) + 2 g \, n(x,g).
\end{equation}

We assume that the semiclassical condition $k_B T \gg \hbar
\omega$ is fulfilled by the system under consideration, so that
the relevant excitation energies are much larger of the level
spacing in the oscillator energies.  In such limit the
single-particle energy in phase space (which is given by the
eigenstates of (\ref{HF Hamiltonian}) ) is

\begin{equation}\label{energy interaction}
E(p,x,g) =  \epsilon(p,x) + 2 g \, n(x,g)
\end{equation}
where $\epsilon(p,x) \equiv p^2/2m + V(x)$, see
\cite{dalfovo,parkinswalls,burnett
}.

Moreover  the semiclassical condition allows approximating
summations over energy states by integrals, namely $\sum
\rightarrow \int d^3xd^3p/h^3$. Therefore, the number of bosons in
the excited spectrum is given by

\begin{equation}\label{semiclassicalNth1}
N - N_0= \int \frac{d^3x d^3p}{(2\pi \hbar)^3}
\left(\exp\left[\frac{E(p,x,g)-\mu}{k_B T}\right]-1\right)^{-1}.
\end{equation}
where $N_0$ is the number of bosons in the condensate and $\mu$
their chemical potential.

When the thermalized gas of bosons reaches the condensation
temperature $T_c$, the chemical potential $\mu$  equals  the
energy of the ground state, i.e.

\begin{equation}\mu_c(g) \equiv min_{x,p}\{E(p,x,g)\} =
min_{x}\{V(x) + 2 g n(x,g)\}.
\end{equation}
We assume that this minimum is reached at $x=0$, so that the
chemical potential at condensation is

\begin{equation} \label{numericalmuc}
\mu_c(g) = 2 g \, n(0,g) .
\end{equation}
We  show later that this assumption is correct, see
Eq.(\ref{semiclassical n 8}) below.

At the condensation temperature the number of bosons in the
condensate $N_0$ is still zero. Thus from
(\ref{semiclassicalNth1}-\ref{numericalmuc}) one has


\begin{equation}\label{numericaltc1}
\begin{array}{ll}
N \pi \hbar^3/2 = \int dx dp x^2 p^2
\left(\exp\left[\frac{E(p,x,g)-\mu_c(g)}{k_B
T_c(g)}\right]-1\right)^{-1}\\
\\
\equiv  \int  d\Omega \,\, \Lambda\left[\theta\right]
\end{array}
\end{equation}
where

\begin{equation}\label{nbardefinition}
\begin{array}{ll}
d\Omega \equiv dx\, dp\,x^2\,p^2; \qquad
\Lambda\left[\theta\right] \equiv
\left[\exp{\left[\theta\right]}-1\right]^{-1};\\
\\
\theta \equiv \frac{\epsilon(p,x)+ 2 \bar{n}(x,g)}{k_B T_c(g)};
\qquad \bar{n}(x,g) \equiv n(x,g)-n(0,g).
\end{array}
\end{equation}
Eq.s(\ref{numericaltc1}-\ref{nbardefinition}) define the
condensation temperature as a function of the parameter $g$. To
obtain the explicit form of $T_c(g)$ one must, in principle,
invert the integral relation (\ref{numericaltc1}). However, we can
avoid exact calculations since, for our purposes, we are
interested only in the shift in the condensation temperature for
small values of the parameter $g$. If $\Delta T_c$ is analytic in
$g$, one can express the relative shift in $T_c$ as

\begin{equation}\label{deltaTsuT0}
\frac{\Delta T_c}{T^0_c} =  \sum_{h=1}^\infty \frac{g^h}{h!}
\frac{\partial_g^h T_c(g)}{T_c(g)}{|_{g=0}}
\end{equation}
where we have used the equality  $T_c(g=0) = T_c^0$. Since in
general one finds

\begin{equation}
\frac{\partial^h_{g}T_{c}(g)}{T_{c}(g)}|_{g=0}=
\frac{I_h}{\left(k_B T^0_c \lambda_{T^0_c}^3\right)^h},
\end{equation}
where the numerical factors $I_h$ can be calculated explicitly,
Eq.(\ref{deltaTsuT0}) can be recast as

\begin{equation}\label{deltaTsuT4}
\frac{\Delta T_c}{T^0_c} =  \sum_{h=1}^\infty \frac{2^h I_h}{h!}
\left(a/\lambda_{T^0_c} \right)^h \equiv \sum_{h=1}^\infty b_h
\left(a/\lambda_{T^0_c} \right)^h
\end{equation}
which defines the coefficients $b_h$. If $\Delta T_c$ is
non-analytic but possesses finite first-$q$ derivatives
$\partial_g\Delta T_c$  at $g=0$, the first $q$ terms in
(\ref{deltaTsuT4}) will give an asymptotic expansion of $\Delta
T_c/T^0_c$ at sufficiently  small $a/\lambda_{T^0_c}$. This is
precisely what happens with the function $T_c(g)$ defined by
Eq.(\ref{numericaltc1}).

Let us  calculate the first derivative $\partial_g T_c/T_c|_{g=0}$
to estimate $b_1$. We note that, since the lhs of
(\ref{numericaltc1}) is independent of $g$ one has $\partial
_{g}N=0$, thus deriving the rhs and equating to zero one has

\begin{equation}\label{derivation1}
\int d\Omega \Lambda'\left[\theta\right]   \partial_g \theta = 0 ,
\end{equation}
with $\Lambda'[\theta] \equiv \partial_\theta \Lambda[\theta] = -
\exp[\theta]/[\exp[\theta]-1]^2$, and after some algebra one
obtains

\begin{equation}\label{deltaTsuT1}
\frac{\partial _{g}T_{c}(g)}{T_{c}(g)}=  \frac{2 \int d\Omega
\left(\bar{n}(x,g)+g \partial_g \bar{n}(x,g) \right)
\Lambda'\left[\theta\right]}{\int d\Omega \left(\epsilon(x,p)+ 2 g
\bar{n}(x,g) \right) \Lambda'\left[\theta\right]}
\label{deltaTsuT1}
\end{equation}
To evaluate (\ref{deltaTsuT1}) at $g=0$ one has to know
$\bar{n}(x,0)$. Since at $T_c$ the number of bosons in the
condensate $N_0$ is zero, one can use (\ref{semiclassicalNth1}) to
express the density of bosons as

\begin{equation}
n (x,g)= (2\pi \hbar)^{-3} \int d^3p\, \Lambda[\theta],
\end{equation}
which gives $N = \int d^3x\,n (x,g)$. This expression can be
integrated to give \cite{dalfovo,parkinswalls,burnett}

\begin{equation}\label{semiclassicaln2}
n(x,g)= \lambda_{T_c}^{-3} g_{3/2}\left[\exp\left[ - \frac{V(x)+2
g \, \bar{n}(x,g)}{k_B T_c(g)}\right] \right],
\end{equation}
where $\lambda_{T_c} = \sqrt{2 \pi \hbar^2 / (m k_B T_c(g))}$ and
$g_{\alpha}[z] = \sum_{k=1}^\infty z^k/k^{\alpha}$ is the
Polylogarithmic or Boltzmann function of index $\alpha$. This is a
consistency relation which can in principle be used to extract
$n(x,g)$. However,  for our purpose we need only $n(x,g)$ (and its
derivatives) evaluated at $g=0$ and not its explicit expression
for any $g$. Since from the last of (\ref{nbardefinition}) it
follows that $\bar{n}(x=0,g)\equiv 0$ for all $g$, Eq.
(\ref{semiclassicaln2}) gives $n(x=0,g) = \lambda_{T_c}^{-3}
g_{3/2}\left[1\right]$ and therefore

\begin{equation}\label{semiclassicaln3}
\bar{n}(x,g)= \lambda_{T_c}^{-3} \,  Q\left[\frac{V(x)+2 g \,
\bar{n}(x,g)}{k_B T_c(g)}\right],
\end{equation}
with

\begin{equation}\label{Qdefinition}
Q[\alpha] \equiv  g_{3/2}\left[\exp\left[ - \alpha\right]
\right]-g_{3/2}\left[1\right] . \end{equation} Then, using
(\ref{semiclassicaln3}-\ref{Qdefinition}) one can evaluate
(\ref{deltaTsuT1}) at $g=0$ obtaining

\begin{equation}
\frac{\partial _{g}T_{c}(g)}{T_{c}(g)}{|_{g=0}}=  \frac{I_1}{k_B
T_c^0 \lambda_{T_c^0}^3} \label{deltaTsuT2}
\end{equation}
where

\begin{equation}\label{I1}
I_1 = 2 \frac{\int d\Sigma \,
\Lambda'\left[u^2+v^2\right]Q\left[v^2\right]}{\int
 d\Sigma \, \left(u^2+v^2\right)
\Lambda'\left[u^2+v^2\right] }\simeq -1.713
\end{equation}
with $d\Sigma \equiv du \, dv \, u^2 v^2$. Therefore, from
(\ref{deltaTsuT4}) one has  $b_1 \simeq -3.426$ \cite{dalfovo}  in
agreement with the experimental value $b_1 = -3.5 \pm 0.3$
obtained in \cite{condensatePRL}.

At that point we can verify  that  the minimum of the energy
(\ref{energy interaction}) is $g \, n(0,g)$, so that
(\ref{numericalmuc}) is  correct. Deriving (\ref{semiclassicaln2})
with respect to $x$ and summing $\partial_x V(x)$ one obtains
\begin{equation}\label{semiclassical n 8}
\partial_x (V+2g n) =  \frac{\partial_x V}{
1 + \frac{2g}{k_B T_c(g) \lambda_{T_c}^3}
g_{1/2}\left[\exp\left[-\frac{V+2 g \bar{n}}{k_B T_c(g)} \right]
\right]}
\end{equation}
and hence $\partial_x (V(x)+2g n(x,g))>0$ for  $x>0$, which in
turn implies that $x=0$ is a minimum of $V(x)+2g n(x,g)$ and
$\mu_c(g) = 2 g n(0,g)$. From integration of (\ref{semiclassical n
8})  one also deduces that close to the center of the trap    at
$V(x)+ 2 g \bar{n}(x,g) \ll k_B T_c$, atoms are quasi-free
particles, since they feel an effective potential $V(x)_{eff}
\equiv V(x)+2g n(x,g) \ll V(x)$.

Let us now estimate the second coefficient $b_2$ in
(\ref{deltaTsuT4}). Deriving (\ref{derivation1}) with respect to
$g$ one has

\begin{equation}
\int d\Omega \{ \Lambda'\left[\theta\right]
\partial_g \theta +\Lambda''\left[\theta\right]
\left(\partial_g \theta\right)^2 \}   = 0 ,
\end{equation}
and after some algebra one obtains

\begin{equation}\label{deltaTsuT5}
\frac{\partial_{g}^2T_{c}(g)}{T_{c}(g)}_{|_{g=0}}=  \frac{\int
d\Omega \,  \left(4 \Lambda'\left[\theta\right] \partial_g \bar{n}
+ \Lambda''\left[\theta\right] \frac{    \left(2 \bar{n}-
\epsilon\frac{\partial_g T_c}{T_c} \right)^2}{k_B T_c}
\right)}{\int d\Omega \, \Lambda'\left[\theta\right] \,
\epsilon}{|_{g=0}}
\end{equation}
where $\Lambda''\left[\theta\right]\equiv
\partial^2_\theta \Lambda\left[\theta\right]$. Since we already know
$\bar{n}(x,g)$ and $\partial_g T_c(g)/T_c(g)$ at $g=0$, the
missing  piece to evaluate (\ref{deltaTsuT5}) is $\partial_g
\bar{n}(x,g)$ at $g=0$. Derivatives of $\bar{n}(x,g)$ can be
obtained by direct derivation of
(\ref{semiclassicaln2}-\ref{semiclassicaln3}). For instance, from
(\ref{semiclassicaln2}) one has  that $n(0,g) = \lambda_{T_c}^{-3}
g_{3/2}\left[1\right]$, and therefore

\begin{equation}
\partial_g n(0,g) =
\frac{3}{2} \frac{\partial_g T_c(g)}{T_c(g)} n(0,g).
\end{equation}
Moreover, from (\ref{semiclassicaln3}) one has

\begin{equation}\label{semiclassicaln6}
\begin{array}{ll}
\partial_g \bar{n}(x,g) = \left(
\frac{3}{2} \frac{\partial_g T_c(g)}{T_c(g)} \bar{n} + \right.\\
\\
\left. + \lambda^{-3}_{T_c}   \left(\frac{V + 2 g \bar{n} }{k_B
T_c}\frac{\partial_g T_c(g)}{T_c(g)}  - \frac{2 \bar{n}}{k_B T_c}
\right) g_{1/2}\left[\exp\left[-\frac{V+2 g
\bar{n}}{k_B T_c(g)} \right] \right] \right)\\
\\
/\left(1 + \frac{2g}{k_B T_c(g) \lambda_{T_c}^3}
g_{1/2}\left[\exp\left[-\frac{V+2 g \bar{n}}{k_B T_c(g)} \right]
\right] \right), \end{array}
\end{equation}
thus

\begin{equation}\label{semiclassicaln7}
\partial_g \bar{n}(x,g)|_{g=0} = \left( k_B
T^0_c\, \lambda_{T^0_c}^6 \right)^{-1}  S\left[-V(x)/k_B
T_c^0\right] ,
\end{equation}
with

\begin{equation}\label{Sdefinition}
S[\alpha] \equiv \frac{3}{2} I_1 Q[\alpha] + \left(\alpha I_1 - 2
Q[\alpha] \right) g_{1/2}\left[\exp\left[-\alpha\right]\right].
\end{equation}
Note that (\ref{semiclassicaln7}) has a finite limit for
$x\rightarrow 0$. In fact, from the expansion

\begin{equation}
g_{s}\left[\exp\left[-\alpha\right]\right] =
\Gamma\left[1-s\right]\alpha^{s-1} + \sum_{k=0}^\infty
\zeta\left[s-k\right] (-\alpha)^k/k!, \end{equation} which is
valid for $|\alpha|< 2\pi$, one finds the asymptotic expansion

\begin{equation}
S\left[-V(x)/k_B T_c\right]\simeq 4\pi + \sqrt{\pi V/k_B
T_c}(4\zeta[1/2]-2 I_1)\end{equation} for small $x$.

Using (\ref{semiclassicaln7}-\ref{Sdefinition}), from
(\ref{deltaTsuT5}) one obtains

\begin{equation}
\frac{\partial _{g}^2 T_{c}(g)}{T_{c}(g)}|_{g=0}=
\frac{I_2}{\left( k_B T_c^0 \lambda_{T^0_c}^3 \right)^2}
\end{equation}
with

\begin{equation}\label{I2}
\begin{array}{ll}
I_{2} = 4 \int \, d\Sigma \, \left[ \Lambda^{\prime
}\left[u^2+v^2\right] S\left[v^2\right] + \Lambda^{\prime
\prime}\left[u^2+v^2\right] \times \right.\\
\\
\left. \left[Q\left[v^2\right]- \frac{u^2+v^2}{2}I_1 \right]^2
\right]/\int d\Sigma \left(u^2+v^2\right) \Lambda^{\prime
}\left[u^2+v^2\right]
\end{array}
\end{equation}
which gives $I_2 \simeq 9.388$ and finally, from
(\ref{deltaTsuT4}) one has

\begin{equation}\label{b2}
b_2 \simeq 18.8 \, .
\end{equation}
We note that this value improves upon the estimation $b_2 \simeq
11.7$ obtained in \cite{condensatePRL2}. Such result was obtained
by numerical methods and therefore it is compatible with
(\ref{b2}), which is an exact analytic result in the MF
approximation, within the precision of the numerical estimation in
\cite{condensatePRL2}. However, (\ref{b2}) is stil not in
agreement with the experimental estimation $b_2^{exp}\simeq 46\pm
5$ \cite{condensatePRL} and this fact can be related to a
beyond-MF effect. Here we propose a different interpretation of
such a disagreement in the framework of MF approximation.

One might  expect that higher order terms in (\ref{deltaTsuT4})
can be important and reduce the difference between $b_2$ and
$b_2^{exp}$. However,  divergences emerge if one tries  to
calculate $b_3$. In fact, the third derivative of
(\ref{numericaltc1}) gives

\begin{equation} \label{numericaltc3}
\int   d\Omega \{ \Lambda'\left[\theta\right]
\partial_g^3 \theta + 3\Lambda''\left[\theta\right]
\partial_g \theta\, \partial_g^2 \theta + \Lambda'''\left[\theta\right]
\left(\partial_g \theta\right)^3\}   = 0
\end{equation}
and  diverges when $g=0$ \footnote{  Note that such divergences
are due to the fact that in the semiclassical limit one assumes
that $\hbar \omega/ k_B T_c \rightarrow 0$ but they do not show if
one maintains $\hbar \omega/ k_B T_c$ finite, see Ref.
\cite{yukalov}.}. Thus one concludes that $\partial_g^3 \Delta
T_c$ diverges in $g=0$ and $\Delta T_c/T_c^0$ must be a
non-analytic function of $a$ as in Eq.(\ref{deltaTsuT Asymptotic})
with $\psi\left[a/\lambda_{T^0_c}\right]$ such that
$\psi\left[0\right] = \psi'\left[0\right] = \psi''\left[0\right] =
0$  and $|\psi'''\left[0\right]| = \infty$, e. g.
$\psi\left[a/\lambda_{T^0_c}\right] \sim
(a/\lambda_{T^0_c})^\sigma$ with $2<\sigma < 3$ or
$\psi\left[a/\lambda_{T^0_c}\right] \sim
g_{7/2}[a/\lambda_{T^0_c}]$.

We stress that the term $\psi\left[a/\lambda_{T^0_c}\right]$ may
be important to explain de discrepancy between
(\ref{deltaTsuNONTUNIFORM1}) and data. To verify this hypothesis
one must find the explicit form of
$\psi\left[a/\lambda_{T^0_c}\right]$ and compare the corresponding
$\Delta T_c/T^0_c$ given by (\ref{deltaTsuT Asymptotic}) with
data. However this seems to be a complicated problem, as difficult
as extracting $T_c(g)$ from (\ref{numericaltc1}), which goes
beyond the intent of this Letter and will be discussed elsewhere.

We have shown that, using the Hatree-Fock and semiclassical
approximations,  the temperature shift $\Delta T_c/T^0_c$ induced
by inter-boson interactions is perturbative to second order in
$a/\lambda_{T^0_c}$ as in (\ref{deltaTsuT Asymptotic}) and non
perturbative to third order. This finding seems to be in contrast
with the result reported in \cite{c1}, where the
interaction-induced temperature shift is estimated as

\begin{equation}
\frac{\Delta T_c}{T^0_c} \simeq c_1 (a/\lambda_{T^0_c}) + \left(
c'_2 + c''_2 \ln[a/\lambda_{T^0_c}]\right) (a/\lambda_{T^0_c})^2
\end{equation}
and is not perturbative at second order in $a/\lambda_{T^0_c}$.
However this result is based on more than the Hartree-Fock and
semiclassical approximations, especially the coefficients $c'_2$
and $c''_2$ have been related to measurements that have been made
in lattice simulations, therefore it is not surprising that it
differs for the result reported in this paper. Moreover the
estimation of the parameters $c_1 \simeq -3.436$, $c'_2 \simeq -32
\pi \zeta[2]/3\zeta[3] \simeq -45.9$ and $c''_2 \simeq -155$ given
in \cite{c1} gives a $\Delta T_c/T^0_c$ which  differs from our
estimation  (\ref{deltaTsuT Asymptotic}) but also from the
estimation (\ref{deltaTsuNONTUNIFORM1}) given in
\cite{condensatePRL2} and from the measurements reported in
\cite{condensatePRL}.

All these considerations show that, before addressing the problem
of beyond-MF effects of interactions in order to explain data in
\cite{condensatePRL}, MF effects should be thoroughly understood.

Before concluding, let us discuss finite size effects on the
condensation temperature $T_c$, which are due to the finiteness of
$N$ in comparison with the thermodynamic limit $N\rightarrow
\infty$ and give a temperature shift $\Delta T^{N}_{c}/T_{c}$. We
note the following: in \cite{condensatePRL} each measurement
series at a given $a$  is compared with a reference measurement at
small $a\simeq 0.005$, with the same $\omega$ and very similar
$N$, so that the measured $\Delta T_c/T_c$ is assumed to be
unaffected from all $a$-independent effects, including systematic
errors in the absolute calibration of $N$ and finite-size effects.
The latter are estimated to leading order  as $\Delta
T^{N}_{c}/T_{c} \simeq -0.73 \, N^{-1/3}$
\cite{grossman,ketterle,kirsten}; this gives $\Delta
T^{N}_{c}/T_{c}\sim 10^{-2}$  for $N\sim 10^5$ as in
\cite{condensatePRL}. Thus finite size effects to leading order
are comparable with interaction effects to leading order
$a/\lambda_{T^0_c}$ but, being independent of $a$, they do not
affect $\Delta T_{c}/T_{c}$ measurements in \cite{condensatePRL}.

It is meaningful to expect that finite size effects to
next-to-leading order also depend on $a$ and give a temperature
shift $\Delta T^{N}_{c}(N,a)$ which, depending on $a$, affects the
$\Delta T_{c}/T_{c}$ measured in \cite{condensatePRL}. Hence,
finite size effects to next-to-leading order may explain the
difference between (\ref{b2}) and $b_2^{exp}$ in the MF framework.

We stress that if this argument is correct and the discrepancy
between (\ref{deltaTsuNONTUNIFORM1}) and data is due to finite
size effects, one expects that in an experiment carried out at
larger N, which minimize finite size effects and  better obeys the
thermodynamic limit, data can be in better agreement with  the
value of $b_2 \simeq 18.8$. However such an experiment is  not
easily realized, since one requires huge $N$s, about three orders
of magnitude above typical values obtained in laboratory BECs, to
significantly diminish finite-size effects. These caveats are
currently under under study and will be presented elsewhere.

In conclusion  we have shown that, under the unique assumptions of
the Hartree-Fock and semiclassical approximations, inter-boson
interactions produce a shift in the condensation temperature of a
trapped BEC which is a non-analytic function of the s-wave
scattering length $a$ at $a = 0$ and it has the asymptotic
behavior (\ref{deltaTsuT Asymptotic}) for weak inter-atom
interactions $a \ll \lambda_{T^0_c}$, so that $T_c(a)$ is
perturbative to second order and non perturbative to third order
in $a/\lambda_{T^0_c}$. We have compared this finding with
different results in \cite{condensatePRL2} and \cite{c1}  and we
have discussed the differences found. We have obtained an
analytical estimation of the parameter $b_2\simeq 18.8$ which
improves previous numerical estimations \cite{condensatePRL2} and
have discussed its remaining discrepancy with atomic BEC
experiments which give $b_2^{exp}\simeq 46\pm5$
\cite{condensatePRL}.

\textbf{Acknowledgements}: FB is a Marie Curie fellow of the
Istituto Nazionale di Alta Matematica. I am very grateful to M. de
Llano, M. Grether, E.M.N. Cirillo and E. Castellanos for useful
discussions during the edition of this manuscript and to B.
Pe\~{n}uela for her strong support.   I also  thank the Instituto
de Investigaciones en Materiales of UNAM University of Mexico City
for the kind hospitality during the edition of this paper and
UNAM-DGAPA-PAPIIT (Mexico) for partial support from grant
IN102011.

\end{document}